\documentclass{article}
\usepackage{graphicx}
\usepackage{amsmath}

\begin{document}

\title{Modelling Epistemic Systems}

\author{Andr\'e C. R. Martins\\
 GRIFE-EACH, Universidade de S\~ao Paulo\\
 S\~ao Paulo, SP, Brazil
}

\maketitle

\begin{abstract}
In this Chapter, I will explore the use of modeling in order to understand how Science works. I will discuss the modeling of scientific communities, providing a general, non-comprehensive overview of existing models, with a focus on the use of the tools of Agent-Based Modeling and Opinion Dynamics. A special attention will be paid to models inspired by a Bayesian formalism of Opinion Dynamics. The objective of this exploration is to better understand 
 the effect that different conditions might have on the reliability of the opinions of a scientific community. We will see that, by using artificial worlds as exploring grounds, we can prevent some epistemological problems with the definition of truth and obtain insights on the conditions that might cause the quest for more reliable knowledge to fail.
\end{abstract}

\section{Introduction}

The classical description of the work of a scientist is one that many of us would like to believe true. According to it, scientists are as close as reasonably possible to believe to selfless individuals who pursue knowledge using always the best means available to them. They propose hypothesis and theories according to what they feel would describe all relevant data better. And, from those models, they draw predictions that are tested under very strict conditions. Finally, still according to our idealized view, it is the agreement between observations and those predictions that dictate which theories are better accepted. Conditions of beauty, like simplicity and symmetry, can be invoked when deciding between equally well suited theories, depending on who is telling the tale. But that is it, and no other considerations should be included.

Obviously, the real life is not so simple and scientists do defend their own ideas because it is their own or because they like it better. And, as in any human activity, one will find individuals whose agenda is based only in self-interest. However, as long as community moves in the right direction and eventually corrects any mistakes, we could, in principle, consider these problems as just perturbations to an otherwise reasonably precise description of the scientific enterprise. In particular, the amazing advances of knowledge in many areas, where old ideas have been replaced by newer and better ones, seems to lend a lot of credit to that idealized description. At the very least, as a good approximation for the system as a whole, even if it fails in individual cases.

And, indeed, we do have controls to avoid the more serious problems. Replication (or reproducibility) \cite{giles06a,palmer06,martins08d}  is considered fundamental by most and with good reason. Once an independent researcher obtains the same results, chances of error in the first report are obviously smaller. But, most important of all, if the groups are truly independent and the second group has nothing to gain from either a positive or a negative result, the chances that the first group published fraudulent results basically disappear. This makes the problem of self-serving scientists much less important than it might otherwise be. It also points out to another central feature of the scientific endeavor that is sometimes neglected. That is, the social aspect of Science.

As in many human activities, the outcomes of scientific research are nowadays a social product of a community of scientists. Theory and measurement are often done by different individuals, except in cases where just a fact, a simple idea, or a property of some material or drug are tested. Even in these cases, there is often a group behind the results, as many fields have become so complex that the expertise of many different researchers is often needed. Main examples of this are the "Big Science" projects, such as the mapping of the human genome or the hunt for the Higgs boson. You have different people taking care of different aspects of the problem, some building the equipments, others operating them, a group specializing in collecting the data, more people to store it properly, others to interpret it and so on. Even in smaller projects, like a simple simulation that one can program alone, we often use software developed by a third party, software we trust to do what we are told it does. We depend on the quality of the information and work performed by others in most of the research done nowadays. And, while it might be reasonable, under some circumstances, to assume that most of the information we use were produced in a competent and honest way, it is certain that we can not claim it for all the information\cite{checkcyranoski05,choetal06}. Despite what we are taught in school, arguments of authority are actually unavoidable \cite{hardwig85a} and we have to understand the effects this might have.

As soon as we realize that we have come to depend on the results of others even to make our own research, we have to realize we might risk running into epistemic trouble. While it is clear that some social aspects are clearly positive (as, obviously, replication), social influence could lead some people away from the best answer. That means that understanding which social aspects can help the reliability of a result and which ones are more likely to be detrimental is an important task. And we must acknowledge that there are different accounts of the scientific process that do not describe it in such beautiful collors, like, per example, the Strong Program in the Sociology of Scientific Knowledge \cite{barnesbloor82a,pickering84a,latourwoolgar86a,bloor91a}. That program defends that Science is just another human social activity. And, as such, not more reliable than any other descriptions of the world obtained by other methods. While scientific advances seem to contradict that in a very strong way, one must admit that some of the criticism might be correct and that there might be ways to make the whole scientific activity even more reliable and less prone to error. In a time where no single human can check the correctness of his whole field alone, understanding the consequences of social interaction and errors can be fundamental to the future quality of the Science we will make. 

Furthermore, there are fields in which published results have been consistently shown to be wrong\cite{ioannidis05b,begleyellis12a,sarewitz12a}, exactly due to the way the Science and publications are structured. Biases towards publishing only some types of results exist and are very prejudicial. It is also already known that replication is a far less common practice than it would be expected in areas as different as Marketing \cite{hubbardarmstrong94a,Evanschitzkyetal07a} and Medicine \cite{ioannidis05a}. These findings make it very clear that it is imperative that we study and understand well the effects of how the way Science is structured and how scientists interact might affect the quality of our conclusions.

Of course, trying to figure how reliable an answer is or if we can really say we know something about the world is an old philosophical problem that we will not answer so simply. However, the question of reliability of results might be particularly well suited to be analyzed with the tools of simulation and agent based models. Simulating a community of scientists will not answer the deepest problems of Skepticism \cite{kleinp02}. However, if we assume that it is possible to know something about the world, 
agent based models of Science can help us explore which strategies and behaviors are more likely to cause problems or to get us closer to the best answers. This happens not only because we can explore unforeseen consequences of interaction rules that we wouldn't see in models built using just human language, but also because we can, as programmers, choose which theory is the best one in the artificial world of our agents. And therefore, in principle, check the effects of different social structures on the choice of the best theory. Of course, any limitations that apply to current models of social interaction will apply here fully and the results must be analyzed with the typical doubt we associate with tentative new ideas. But we can get some advice on where it might be more likely that a change in current practices will lead to better outcomes.

\subsection{Existing models for describing scientific practices}

In this paper, I defend one specific use for agent models, one that has seen a small number of papers up to now. However, it is interesting to acknowledge that the applications of computational and mathematical modeling to communities of scientists is an area that has been seen a steady increase in the interest and number of papers. Different researchers have dedicated themselves to different problems, most of them tending towards describing the society of scientists. Therefore, it makes sense to provide a very brief and very incomplete review of those applications. Those works can be divided in three major areas: methods to measure the quality and impact of a scientist work; studies of the structure of Science, as per example, the networks of authors and their papers; and the modeling of knowledge, either just looking at its difusion, or at the acceptance of new and better knowledge. It is particularly worth noting that recently, there was a full edition of the Journal of Artificial Societies And Social Simulation \cite{edmondsetal11} as well as a book \cite{scharnhostetal12a}, both dedicated to simulation of Science.

Measuring the impact of the work of each of us can be an important task. It has significant influence on our careers and scientific policy makers could use such this information when deciding how to divide the funding \cite{yilmaz11a}. It is no surprise therefore that attempts at providing quantitative measurements for the relevance of a scientist work have become popular and several different proposals now exist. They include the now unavoidable (despite the biases it introduces) H-index \cite{hirsch05a}, initially proposed to avoid the problems of measurement brought by total number of citations, that could be a very large number even if the scientist had just collaborated in one very well cited article. Several other measurements were also proposed \cite{egghe06a,galam11a} and many others are probably being investigated. One interesting aspect of the problem that still needs to be addressed, however, is the impact these indexes have in scientific production. Scientists are usually reasonably intelligent individuals and, therefore, they might change the way they work in order to have better values at the measurement they are evaluated with, instead of worrying only about making better Science.

On the evolution of scientific fields, the ability to measure and quantify production can be both important and interesting in itself \cite{eggherousseau90a}. The  existence of big databases of published material has allowed a large number of studies on structure that is formed by articles and researchers. There are models describing the networks that are formed when scientists cite others \cite{garfieldetal64a,borneretal04a,radicchietal09a,quattrociocchiamblard11a,renetal12a} as well as the networks of co-authorship  \cite{newman01a,newman01b,barabasietal02a}, and even of articles connected by the same specific area \cite{herreraetal10a}. This mapping effort allows us to have a much clearer view of how Science is done. Also, the existence of these studies and the data they are based on provides us with indirect measures of social influences among researchers. A complete agent model of scientific activities should either use or explain the structures that were observed; however, we are still far from obtaining such a model, the same way that we are far from having a really reliable and complete model of how people influence each other choices in any field of human behavior. Existing models, as we will see, are still simplistic versions built to address just a few questions and not really general enough.

Still, more recent years have seen the appearance of the first models of idea diffusion in scientific communities, including approaches that make use of the Master equation \cite{kondratiuketal11a} and population dynamics \cite{vitanovausloos12a} instead of agent models. Among the agent models, several just try to describe the statistical features observed in the studies that measured number of papers and how they are connected by proposing simple rules of influence \cite{gilbert97a,simkinroy05a,simkinroy07a,newman09a} such as, per example, copying the citations of read papers. Attempts at introducing more sophisticated agents, with better cognition, also exist \cite{navehsun06a}. Many general ideas not completely developed have also been proposed as possible basis for future models \cite{doran11a,parinovneylon11a,payette11a}, including general ideas about the practice inspired by the works of philosophers of Science\cite{edmonds11a}.

Of particular interest to the view defended here are models for the diffusion of ideas and opinions that try to answer whether our current practices are good enough or whether we should work to change some of them. One first model used the Bounded Confidence model \cite{deffuantetal02a,hegselmannkrause02} of Opinion Dynamics \cite{castellanoetal07,galametal82,sznajd00,hegselmannkrause02,galam05b,martins08a,martins12a} as basis, exploring if agents would approach a true value of a parameter given that some of the agents were truth-seekers, who had a tendency to move towards the correct result \cite{hegselmannkrause06a}. However, by adopting a purely continuous version of opinion, that model, while very interesting, is not enough to describe the choice of theories. The effects of peer review \cite{sobkowicz10a,squazzonitakacs11a} were also suggested as an important field to explore by the use of models, since many of the proposals for change in the peer review process have not been explored wit the necessary depth. In the same line, discussions of the possible use of modelling to make Science more reliable already exist \cite{zollman11a}. However, a strong push forward in developing these models is still to be observed. 

Another promissing line of investigation on the reliability of Science is the study of the effects of the existence of agents who make claims that are stronger than the actual observations warrant. The main question explored there is if this type of lie could help in the convincing of general public\cite{galam06b,galam10a}. Finally, I have also proposed a model to explore acceptance of better theories \cite{martins10a}, one that was able to show that the idea that Science advances thanks to the retirement of old scientists \cite{kuhn} might be true. In the Section \ref{sec:codamodel}, we will discuss it  and I will show why I believe it can provide the basis for a more sophisticated and maybe even a little less artificial description of the scientific enterprise. That study was based on an Opinion Dynamics model that distinguishes choices and strength of opinion\cite{martins08a,martins08b} and is also based on notions of rationality. A model working on the same problem, where actual publications were introduced as the means that scientists use to communicate their results was also recently proposed with some very promissing results \cite{sobkowicz11a}.

\section{Epistemology and Modelling}

Before presenting a model, it makes sense to discuss a few issues regarding the aplicability of mathematical and computational modeling to the epistemic problem of finding better ways to obtain reliable knowledge. The first question worth mentioning, but too complex to be discussed properly here, is exactly the meaning of knowing \cite{shope02a}. Traditional accounts define that one knows something when the person has a justified true believe about that something, but there are known problems with that definition \cite{gettier63}.

As mentioned above, one interesting feature of artificial worlds is that, since the programmer is playing the part of a creator god, we can circumvent, for the agents, any of the arguments of Skepticism about the possibility of knowing something about the world. Even softer skeptical arguments, like the problem of induction \cite{hume,howson03} can be either dismissed by constructing a world where the rules don't change, or, eventually, one could investigate what would happen if the future did not exactly follow the same laws as the past. By controlling the laws of the artificial universe, different scenarios and their impact on the learning of the agents can be much better investigated than in the real world, where we can only know our theories, but not be always certain about which of them are really better. Of course, this means developping better models for the dynamics of scientific opinions, but the possibility that those models can be quite helpful at exploring epistemic problems should be clear. 

In the next Section, I will describe a model that I believe has the potential to be built upon so that we can arrive at a more realistic description of the problem. But, before that, it is good to debate the fundamental ideas behind it and how they relate to current developments in Epistemology. One line of reasoning in Epistemology claims that all we can really know about the world are probabilities that our ideas are correct. When making any new observation, those probabilities must change according to what we have learn, by obeying the Bayes Theorem. This idea can be called Bayesian Epistemology \cite{kaplan96a,kaplan02,bovenshartmann,talbott08a}. This is a normative theory, in the sense that it claims that this is the best way to make inferences, and not necessarily the way we actually perform them. The claim that scientist behave and choose the theories they support in a way similar to the specifications of Bayesian methods is known in Philosophy as Confirmation Theory \cite{earman92,jeffrey83,jeffrey04,howsonurbach,maher93,bovenshartmann}. While it is known that humans are actually bad probabilists \cite{plous93}, there is some evidence that scientists do not deviate so much from a Bayesian point of view \cite{presstanur01a}. Interestingly, not only scientists, but  there is a growing body of evidence that the way people reason about problems and theories is actually close to Bayesian analysis \cite{tenenbaumetal07,kempetal10a}, if not exactly so. The so-called error in human probabilistic reasoning can actually be explained by assuming that our brains work with more complex models of reality than what was supposed in the  laboratory tests \cite{martins05b,martins06}. 

Therefore, we can use Bayesian models as basis for the behavior of our agents \cite{martins08e}, even if only as a reasonable first aproximation. Such an approach has already produced a number of Opinion Dynamics models \cite{banerjee92,orleans95,lane97,martins08a,martins08b,martins08c,vicenteetal08b,martinspereira08a,sietal10a,martins10b,martins10a,martinskuba09a,martins12a} and it was also used to show why a non-specialized reader should only come to trust a scientific model in cases where there might be errors and deception from the part of the authors, when the results are replicated by a third party \cite{martins05,palmer06,martins08d}.

\section{A Model for the spread of a new theory}\label{sec:codamodel}

Assume each agent must choose between two theories, $A$ and $B$. This choice is a discrete choice between those alternatives, but agents can have a strenght of opinion, represented by the probability  $p_i$ that each artificial scientist $i$ assigns to the possibility that $A$ is the best description. If $p_i>0.5$, agent $i$ believes $A$ is the best option, otherwise, it is $B$. This model is basically the ``Continuous Opinions and Discrete Actions'' model (CODA model), previously used to explore the emergence of extremism in artificial societies \cite{martins08a,martins08b}. CODA is based on the idea that agents believe that, if $A$ is the best choice, each of the neighbors of an agent, located in a given social network \cite{newmanetal06a,vegaredondo2007a}, will have a probability $\alpha>0.5$ of choosing $A$(and similarly, for $B$). In this context, it is easier to work with a transformation of variables, given by the quantity
\[
\nu_i=\ln\frac{p_i}{1-p_i}.
\]
Here, if $\nu_i>0$, we have $p_i>0.5$ and, therefore, a subjective belief in favor of $A$; if $\nu_i<0$, the agent chooses $B$. By applying Bayes Theorem, we obtain a very simple update rule for $\nu_i$, when agent $i$ observes the choice of its neighbor $j$, given by
\[
\nu_i (t+1)= \nu_i (t)+ \textnormal{sign}(\nu_j)*a,
\]
where $a$ is a step size that depends on how likely the agents believe it is that their neighbors will be correct, that is, it is a function of $\alpha$. If we renormalize the update rule, by using $\nu_i^*=\nu_i /a$ instead, we will have
\begin{equation}
\nu_i^* (t+1)= \nu_i^* (t)+ \textnormal{sign}(\nu_j^*),
\end{equation}
making it clear that the value of $a$ is irrelevant to the dynamics of choices, since that dynamics depends only on the signs of $\nu_i$ (or $\nu_i^*$).  The update of the opinions is asynchronous. 

In order to introduce the influence of observations in this problem, a proportion $\tau$ of the scientists actually perform experiments. This experiments can have a stronger or weaker influence on this agents, measured by the probability $\rho$ and experimenter will, at each interaction, observe the Nature, instead of being influenced by a neighbor. In here, we assume experiments always provide the same answer, agreeing with $A$ and that the agenbt will update its opinion by the same amount it would from an interaction with a neighbor. Notice that a stronger effect can easily be introduced simply by allowing $\rho$ to be larger and that means no new parameter needs to be introduced here.

It was observed that, unless $\tau$ is reasonably large, even when experimentalists are very weakly influenced by social effects ($\rho$ close to 1.0), they fail to convince the whole population and clusters of agents preferring the worst theory survive. This effect was particularly strong when theory $A$ was new and, therefore, started with a small proportion of supporters, indicating that the old view would survive simply from social effects and the strength of opinion of its old supporters. Interestingly, when retirement was introduced in the system, with agents replaced by new ones with moderate initial opinions, it became much easier for new and better ideas to invade, confirming Kuhn's notion that Science would advance and accept new paradigms due to the death of the old scientists \cite{kuhn}.

\begin{figure}[htp]
\hspace{-0.5cm}\includegraphics[width=0.32\textwidth]{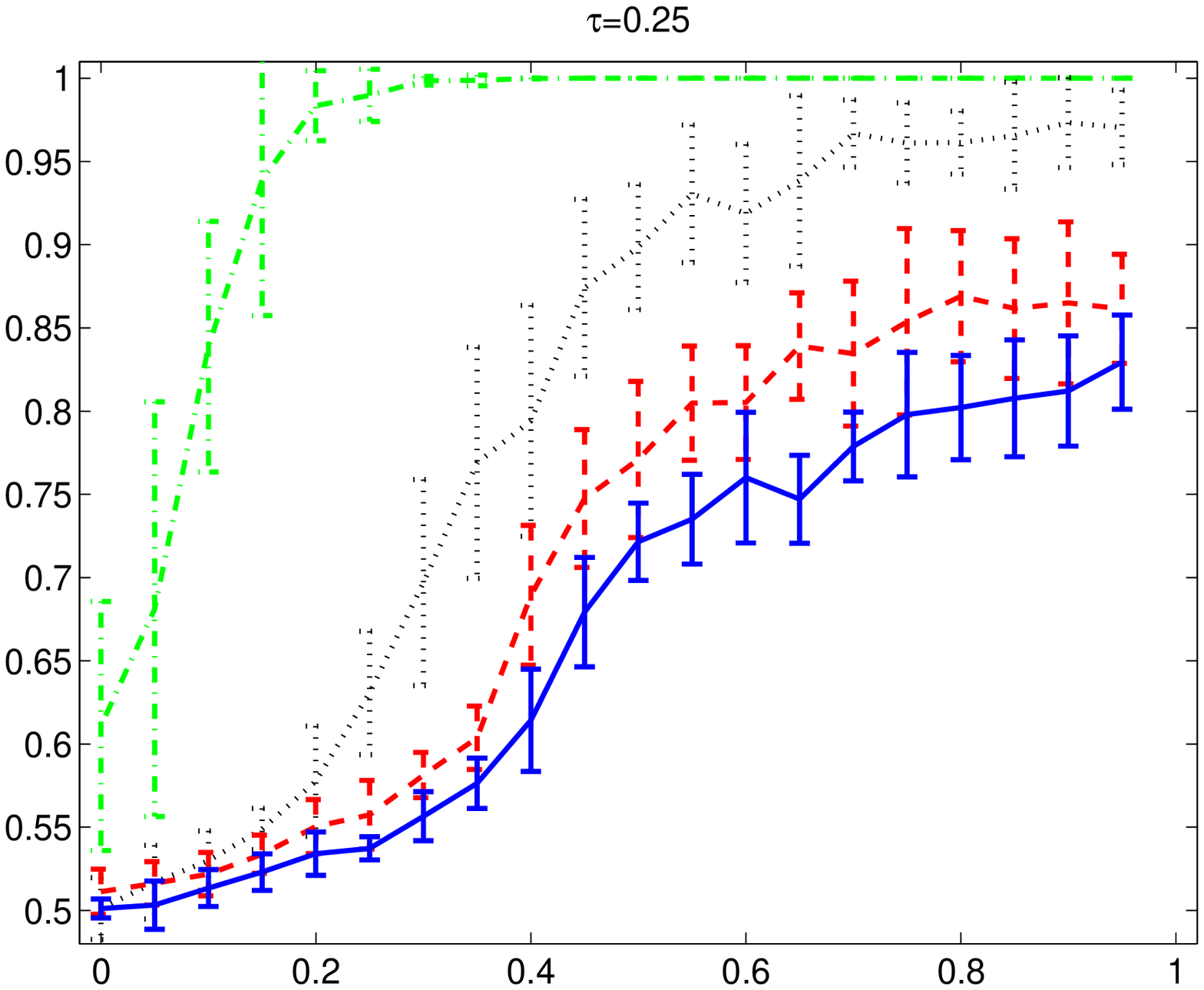}\hspace{0.4cm}
\includegraphics[width=0.32\textwidth]{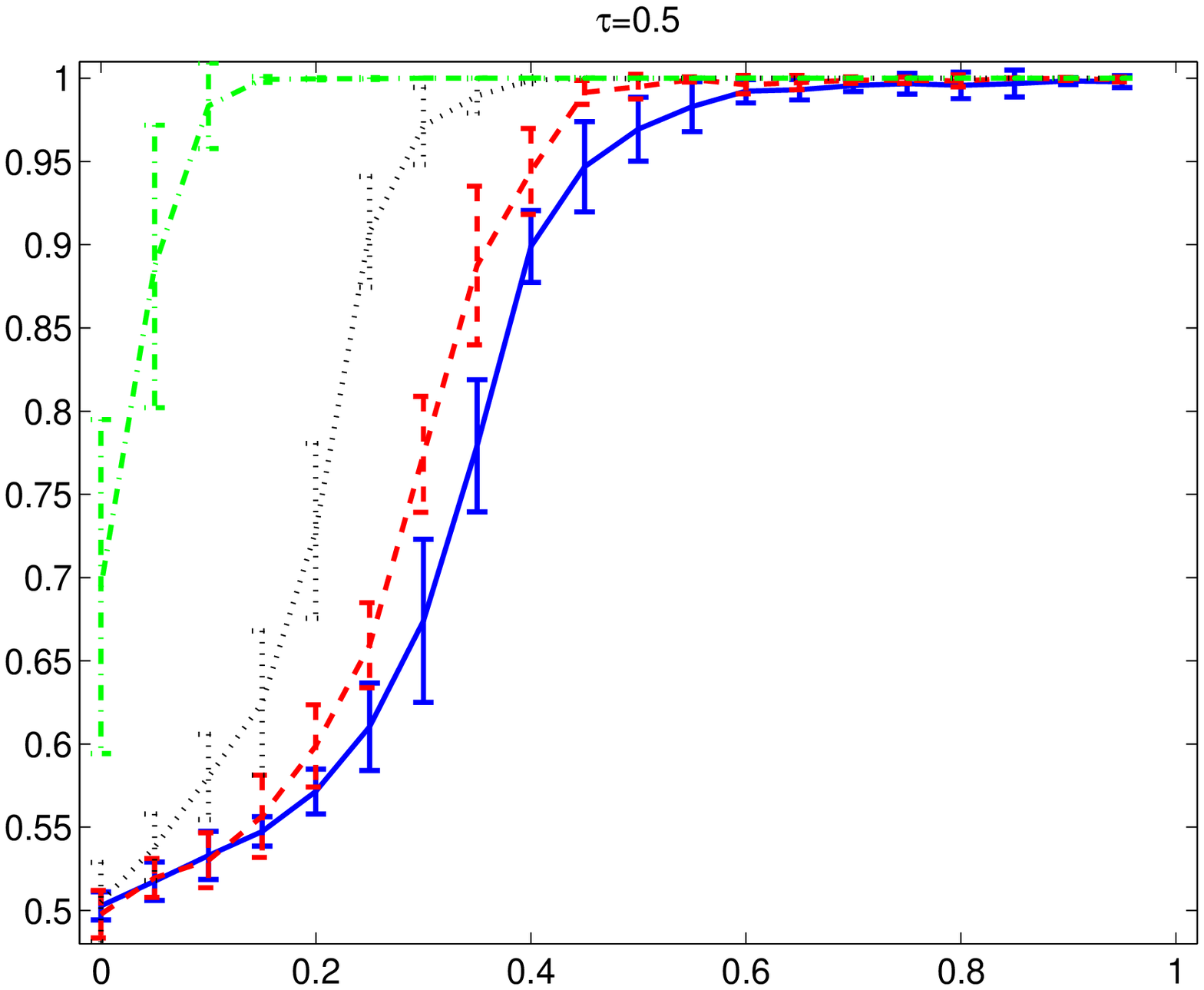}\hspace{0.4cm}
\includegraphics[width=0.32\textwidth]{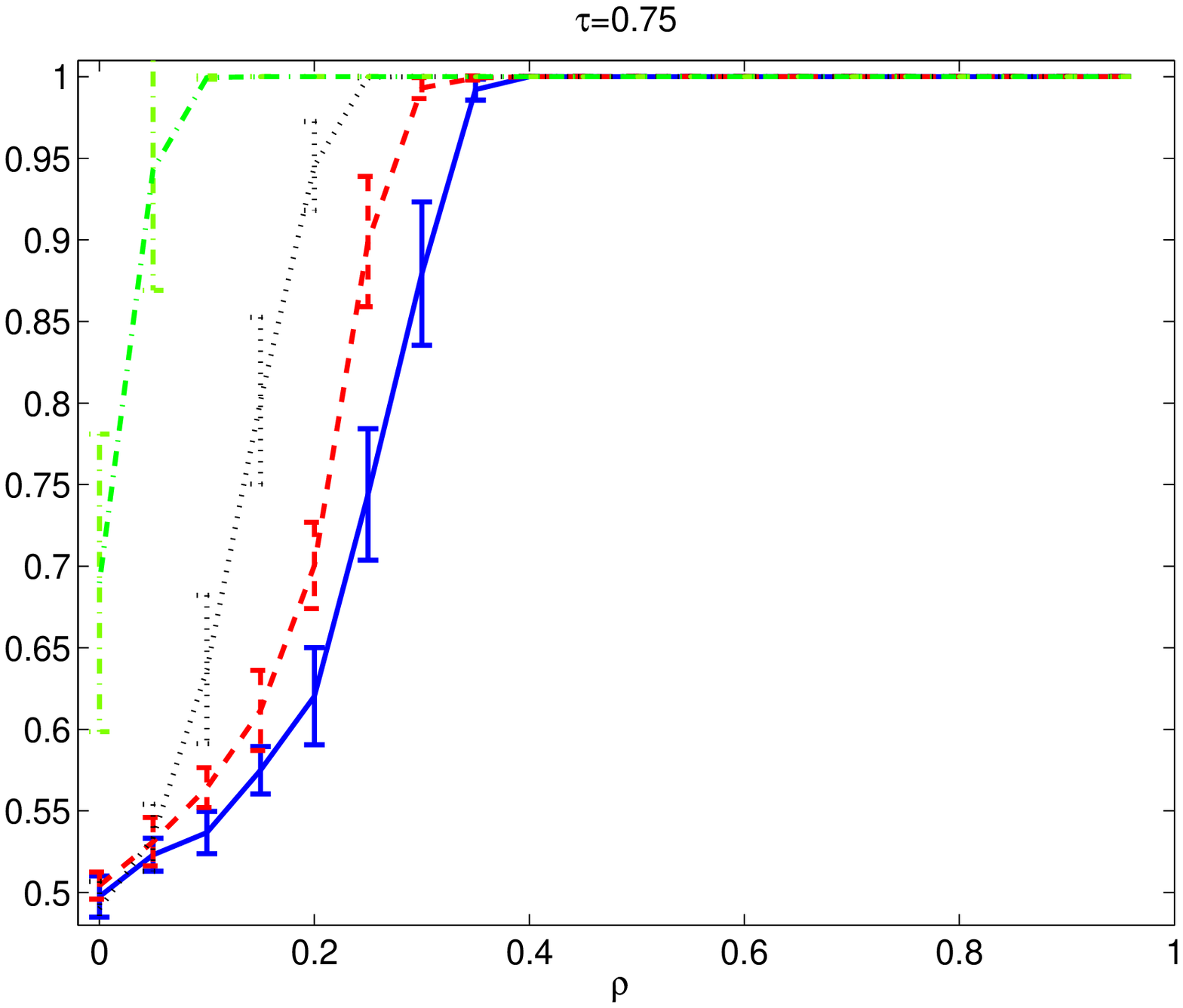}
 \caption{Proportion of agents aligned with the external field (correct scientists) as a function of $\rho$.  Every realization had initial conditions where each agent had 50\% chance of supporting either theory. The solid line corresponds to $\lambda=0.0$, the dashed one to $\lambda=0.1$, the dotted one to $\lambda=0.25$, and the dash-dotted one to $\lambda=0.5$.                                                      
 (\textit{Left panel}) $\tau=25\%$.  (\textit{Middle panel}) $\tau=50\%$. (\textit{Right panel}) $\tau=75\%$.  }\label{fig:sw}
 \end{figure}

Of course, those results were obtained with a very simplified version of the scientific activity. This was intentional, in the old Physics tradition of starting at the very simplest model one can imagine and later add new features, as they become needed. The idea is to understand which details are responsible for which features of the  whole problem. As such, expanding the model from its several approximations is a natural next step. Per example, in the model, social influence happened only between peers and they were all located on a non-realistic square lattice. No publications existed in the model and Nature always gave the same answer, meaning one theory was clearly superior to the other. And, while people would reinforce their ideas given a certain neighborhood, all agents were supposed to be honest, in the sense that they would always choose the theory they really believed more likely to be truth. Despite all this, the model is very easy to expand. By changing the likelihoods agents assign to the choices of others, we could introduce more realistic versions of social influence. Researchers could do papers, adding a new layer to the problem, and the quality of those papers could depend on the correctness of their argument. And, of course, more realistic networks can be trivially introduced. 

In order to illustrate these differences, new cases for small-world networks were run specifically for this paper. The results are shown in Figure \ref{fig:sw}. All cases shown correspond to a initial square lattice with $32^2$ agents, where each link was randomly rewired with probability $\lambda$. Each point is the result of the average of 10 diferent realizations of the same problem and the bars correspond to the standard deviation of the observed values. Initial conditions were such that at first both theories had equal amount of support and the system was left to interact for $t=200$ average interactions per agent. Each panel shows the results for different values of $\lambda$ for a specific proportion $\tau$ of experimenters. We can see that as the small-world effect becomes more important (larger rewiring $\lambda$), it becomes easier for the system to reach agreement on the best theory. However, the problem with the continuing existence of groups supporting the worse theory remains for the cases where the importance of experiment is not large enough, confirming, at least, qualitatively, the results previously obtained.

\section{Conclusion}

We have seen that mathematical and computational modeling of scientific agents can be used as a tool to help us understand under which conditions scientific results are more reliable. While it is quite true that the models so far proposed are very simple, we have seen that they can already capture some of the features of the scientific enterprise, such as the need for solid and reliable experimental data. If experiments are not very convincing, we would be in the situation where $\rho$ is small and, therefore, social influence can indeed become the major force in the system, as suggested in the Strong Prgram. However, as soon as experiments become more important and we have a better connected world, the case for the better theory becomes strong enough to convince everyone. We should also notice that the same model was able to show the importance of the retirement of scientists, as a way to allow new better ideas to spread more easily.

As such, I would like to encourage the modeling community to investigate models for scientific activity. And by that I mean not only from the sociological (and also interesting) point of view of measuring our activities, but also from the point of view of suggesting better practices. While doubt will remain when a suggestion comes from a model, the tools we have make the exploration of the consequences more reliable than simple human intuition and spoken arguments. Models of Science can and should help us make research an even better tool for understanding the world.

\section{Acknowledgements}
The author would like to thank  Funda\c{c}\~ao de Amparo \`a  Pesquisa do Estado de S\~aoPaulo (FAPESP),  under grant 2011/19496-0, for the support to this work.

\bibliographystyle{plain}
\bibliography{biblio}

\end{document}